\documentclass[%
 reprint,
 amsmath,amssymb,
 aps,
]{revtex4-1}

\usepackage{graphicx}% Include figure files
\usepackage{dcolumn}% Align table columns on decimal point
\usepackage{bm}% bold math
\usepackage{amsmath}
\usepackage{amssymb}
\usepackage{latexsym}
\usepackage{epsfig}
\usepackage{amsbsy}
\usepackage{array}
\usepackage{amssymb}
\usepackage{setspace}

\begin{document}

\preprint{}

\title{Topological Weyl and Node-Line Semimetals in Ferromagnetic Vanadium-Phosphorous-Oxide $\beta$-V$_2$OPO$_4$ Compound  }% Force line breaks with \\

\author{Y. J. Jin$^1$$^{,*}$, R. Wang$^{1, 2}$$^{,*}$, J. Z. Zhao$^{1,3}$, Z. J. Chen$^{1,4}$, Y. J. Zhao$^{4}$, and H. Xu$^{1}$$^{,\dag}$}
\affiliation{$^1$Department of Physics, South University of Science and Technology of China,
Shenzhen 518055, P. R. China.}
\affiliation{$^2$Institute for Structure and Function $\&$
Department of physics, Chongqing University, Chongqing 400044, P. R. China.}
\affiliation{$^3$Dalian Institute of Chemical Physics,Chinese Academy of Sciences, 116023 Dalian, P. R. China}
\affiliation{$^4$Department of Physics, South China University of Technology, Guangzhou 510640, P. R. China}

\begin{abstract}
%\baselineskip 12 pt

%We propose the non-trivial ferromagnetic ground state of vanadium-phosphorous-oxide $\beta$-V$_2$OPO$_4$ compound from first-principles. In this magnetic system, with crystal inversion symmetry, the direction of spin-polarization could switch the band structures from a node-line type to a Weyl one in the presence of spin-orbital-coupling. The node-line loop is symmetric protected when the spin is only polarized along [001] direction and will degenerate to only one pair of largely separated Weyl points when the direction of spin-polarization is deviated from [001]. This Weyl semimetal state provides a nice candidate with the minimum number of Weyl points in a condensed matter system. The results of surface band calculations confirm the non-trivial topology of the proposed compound. This findings provides a realistic candidate for the investigation of topological semimetals in the absence of time-reversal symmetry, particularly towards the realization of quantum anomalous Hall effect in Weyl semimetals.

We propose that the topological semimetal features can co-exist with ferromagnetic ground state in vanadium-phosphorous-oxide $\beta$-V$_2$OPO$_4$ compound from first-principles calculations. In this magnetic system with inversion symmetry, the direction of magnetization is able to manipulate the symmetric protected band structures from a node-line type to a Weyl one in the presence of spin-orbital-coupling. The node-line semimetal phase is protected by the mirror symmetry with the reflection-invariant plane perpendicular to magnetic order. Within mirror symmetry breaking due to the magnetization along other directions,  the gapless node-line loop will degenerate to only one pair of Weyl points protected by the rotational symmetry along the magnetic axis, which are largely separated in momentum space. Such Weyl semimetal phase provides a nice candidate with the minimum number of Weyl points in a condensed matter system. The results of surface band calculations confirm the non-trivial topology of this proposed compound. This findings provide a realistic candidate for the investigation of topological semimetals with time-reversal symmetry breaking, particularly towards the realization of quantum anomalous Hall effect in Weyl semimetals.

%Based on first calculations, we theoretically propose that the Weyl semimetal features can co-exist with the ferromagnetic ground state in vanadium-phosphorous-oxide $\beta$-V$_2$OPO$_4$ compound. We find that, so the current compound  . The Weyl points arise from the nodal ring gapped due to the mirror reflection symmetry broken in the presence of spin-orbital-coupling, and are protected by the twofold rotational symmetry along the magnetization direction. Surface band calculations reveal non-trivial Fermi arcs states as well.
\end{abstract}

%\pacs{73.20.At, 71.55.Ak, 74.43.-f}

%Use showkeys class option if keyword %display desired

\maketitle
\baselineskip 12 pt

Quantum-Hall-effect (QHE) offered us a different pathway to achieve dissipationless current beyond superconductors \cite{Hall1879}. However, the potential applications of QHE are strongly limited by the required external high-intensity magnetic fields \cite{Klitzing1980,Tsui1982}. An alternative option is the Quantum-spin-Hall (QSH) insulators that works independently with respect to external magnetic fields but, unfortunately, will be suppressed when the sample size become larger than a critical value due to the existence of non-elastic scattering between states with opposite directions in helical edge states  \cite{Murakami2003,Bernevig2006,Konig2007}. The most promising solution of these challenges is the proposed quantum-anomalous-Hall (QAH) materials with intrinsic long-rang magnetic order that are able to be free for any sample-size or external field \cite{Qi2006,Yu2010,Chang2013}.

To achieve this, magnetic topological materials, including topological insulators (TIs) \cite{Qi2006,Yu2010,Chang2013,Fang2014,Chang2015} and topological semimetals (TSMs)\cite{Xu2011}
in their quantum-well structure with time-reversal (TR) symmetry breaking, have attracted intensive attention recently. Distinguished from gaped magnetic TIs, magnetic TSMs,
i.e. spin polarized Weyl and node-line semimetals, host finite numbers (Weyl) or continues distributed (node-line) band crossing points near the Fermi level in momentum space.
 Such crystal symmetry protected features lead exotic conductive surface states, which are Fermi arcs in Wely semimetals (WSMs) \cite{Wan2011} and drumhead states in node-line semimetals (NLSs) \cite{Burkov2011}. Accomplishing these robust spin dependent surface states is the key step of achieving several unusual spectroscopic and transport phenomena especially the QAH in their quantum-well structure experimentally.

%Recently, topological semimetals (TSMs) have been attracted considerable attention since they extend the topological classification of mater beyond the insulators and exhibit the exotic surface state \cite{Wan2011, Xu2011}. In these materials, the valence and conduction bands disperse linearly around special points in three-dimensional (3D) momentum space, called Weyl points, which construct the discrete point-like Fermi surface. The Weyl point behaves as a monopole and can be quantified by the corresponding chiral charge through calculating the flux of Berry curvature \cite{Wan2011}. Due to the conservation of chirality, the Weyl points always appear in pairs of opposite chirality. The topological Fermi arcs are arising from the connection of two projections of the bulk Weyl points with opposite chiral charges in the surface Brillouin zone (BZ). WSM and its surface state may lead to unusual spectroscopic and transport phenomena such as chiral anomaly, spin and anomalous Hall effects \cite{Xu2011,Adler1969, Shekhar2015, Sun2016}. During the past few years, WSMs were intensively studied theoretically \cite{Wan2011, Xu2011,Weng2015,Huang2015,Ruan2016nc,Ruan2016prl,Autes2016,Wang2016prl1,Wang2016prl2,Xu2017prl}, and the surface Fermi arc states were also experimentally observed in the TaAs family \cite{Xu2015,Lv2015,Lv2015} and MoTe$_2$ \cite{Tamai2016} using angle-resolved photoemission spectra(ARPES).

Up to now, the non-magnetic TSMs have been studied intensively \cite{Weng2015,Huang2015,Ruan2016nc,Ruan2016prl,Autes2016,Wang2016prl2,Xu2017prl,Xu2015,Lv2015,Yang2015,Tamai2016},  but only a few magnetic TSMs have been proposed. For instance, magnetic HgCr$_2$Se$_4$ has been predicted to host only two nodal points \cite{Xu2011} which is defined as a Chern semimetal, not fully fit the WSM features since each crossing points possessing chiral charge of 2. So far, nontrivial properties in HgCr$_2$Se$_4$ have not been verified experimentally due to the requirement of large magnetic domains \cite{Liu2014,Bulmash2014}.  Very recently, Co-based magnetic Heusler alloys were also predicted to host the topology of WSMs \cite{Wang2016prl1}. However, the energy of Weyl points is much higher above the Fermi level ($\sim$0.6 eV). Therefore, additional tuning of the energy of Weyl points relative to Fermi level is necessary \cite{Wang2016prl1} for future applications.

%Weyl points are twofold degeneracy and only exist in condensed matter systems with breaking either time-reversal or spatial-inversion symmetry. For nonmagnetic WSMs by breaking spatial-inversion symmetry, most of them have large number of Weyl points, such as TaAs-family, in which there are 24 Weyl points giving rise to rather complicated transport properties \cite{Weng2015}. In comparison, the ferromagnetic WSMs with time-reversal-breaking may have less Weyl points due to the fermion doubling problem \cite{Wan2011, Xu2011, Wang2016prl1}. For instance, magnetic HgCr$_2$Se$_4$ has been predicted to host only two nodal points \cite{Xu2011}. This compound is a Chern (not Weyl) semimetal due to each crossing points possessing chiral charge of 2. So far, nontrivial properties in HgCr$_2$Se$_4$ have not been verified experimentally since alignment of large magnetic domains is required \cite{Liu2014,Bulmash2014}.  Very recently, Co-based magnetic Heusler alloys were also predicted to host Weyl points with large separation in momentum space \cite{Wang2016prl1}. Although these materials have high Curie temperatures up to room temperature, the energy of Weyl points are about $\sim$0.6 eV above the Fermi level. Therefore, to observe the topological phenomena induced by Weyl fermions, fine tuning of the energy of Weyl points relative to Fermi level is necessary \cite{Wang2016prl1}.

In this work, we propose that the ferromagnetic (FM) vanadium-phosphorous-oxide $\beta$-V$_2$OPO$_4$, a potential material for lithium-ion battery \cite{Benser2007,Satyanarayana2017}, presents either NL or WSM features that can be easily switched by different magnetization directions.  The node-line band structures, in $D_{4h}$($C_{4h}$) magnetic group with spin-orbital coupling (SOC), are protected by the mirror reflection symmetry with the reflection-invariant plane perpendicular to the spin-polarized direction. This gapless node-line loop, which lies in the reflection-invariant plane and very close to the Fermi level, will degenerate to two large separated Weyl points protected by rotational symmetry in the same plane once mirror reflection symmetry is broken due to the magnetization along other directions. In these cases, the system host the minimum number of Weyl points. Based on our first-principles calculations, this non-trivial FM $\beta$-V$_2$OPO$_4$ is a promising candidate of experimentally studying on magnetic TSMs.

%In this letter, we propose that WSM features can co-exist with the ferromagnetic (FM) ground state in vanadium-phosphorous-oxide $\beta$-V$_2$OPO$_4$. This compound, in which there are only two Weyl points very close to Fermi level, possesses the minimum number of Weyl points in a condensed matter system for the first time. Interestingly, along a special magnetization direction, $\beta$-V$_2$OPO$_4$ compound can host the node-line semimetallic features arising from the mirror reflection symmetry even though the spin-orbital coupling (SOC) is included.  $\beta$-V$_2$OPO$_4$ was also of interest because of potential applications for lithium-ion battery \cite{Benser2007,Satyanarayana2017}. Based on our first-principles calculations, it is demonstrated that $\beta$-V$_2$OPO$_4$ is a FM half-metal and will be great use for applications in spintronics. More importantly, the topologically semimetallic features with quantized Berry phase at Fermi level would provide a promising platform to study quantum anomalous Hall effect in experiments.

We perform the first-principles calculations using the Vienna \textit{ab initio} Simulation Package (VASP) \cite{Kresse2,Kressecom} based on density functional theory \cite{Hohenberg,Kohn}. For the exchange-correlation potential we choose the generalized gradient approximation (GGA) with the Perdew-Burke-Ernzerhof (PBE) formalism \cite{Perdew1,Perdew2}. The core-valence interactions are treated by the projector augmented wave (PAW) method \cite{Blochl,Kresse4,Ceperley1980}. A plane-wave-basis set with kinetic-energy cutoff of 600 eV has been used. The full Brillouin zone(BZ) is sampled by $21\times21\times21$ Monkhorst-Pack grid to simulate the electronic behaviors \cite{Monkhorst}. Due to the strongly correlated effects of $3d$ electrons in vanadium, GGA+U calculations is nessesary to describe the on-site Coulomb repulsion beyond the GGA pictures \cite{Liechtenstein1995,Korotin1998}. We notice that the band inversion has been confirmed in a large range of U in $\beta$-V$_2$OPO$_4$ from 1.5 eV to 10.0 eV. In this work, the effective on-site Coulomb energy U is chosen to be 3.0 eV to illustrate the band topology, which works well in fitting the properties of vanadium-oxide system \cite{Liebsch2005}.  To calculate the surface states and Fermi arcs, the tight-binding Hamiltonian is constructed by projecting the Bloch states into maximally localized Wannier functions \cite{Marzari2012,Mostofi2008}.

\begin{figure}
	\centering
	\includegraphics[scale=0.034]{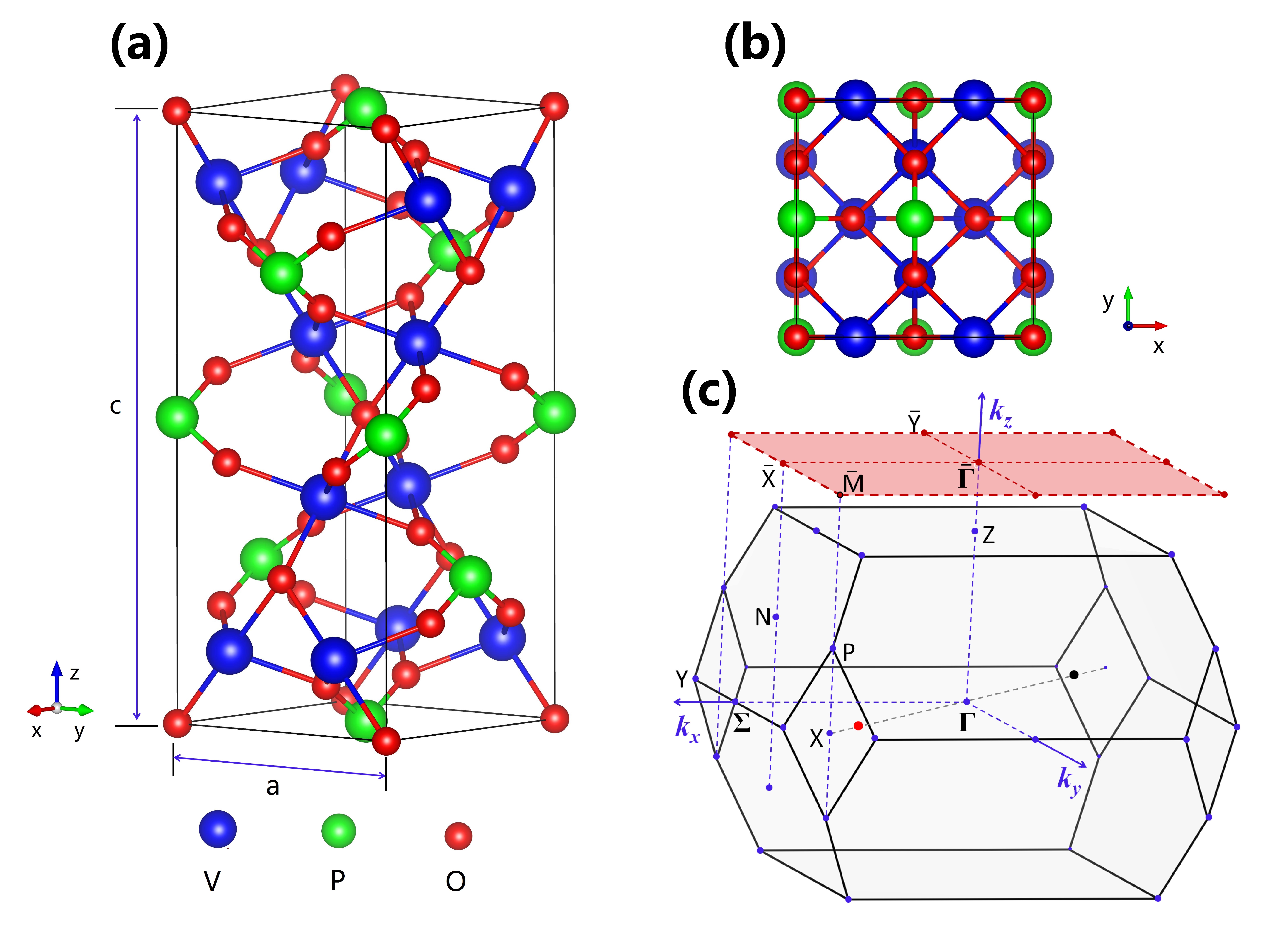}
	\caption{ Crystal structure and Brillouin zone (BZ).  The side (a) top (b) view of crystal structure of $\beta$-V$_2$OPO$_4$  with the space group $I4_{1}amd$ (No. 141). V, P, and O atoms are indicated by blue, green, and red spheres, respectively. (c) The body-centered-tetragonal (BCT) BZ and the corresponding (001) surface BZ. The high symmetry points are also indicated. Two Weyl points with opposite chirality are symmetrically located at $\Gamma$-X axis as red ($C=+1$) and black ($C=-1$) dots, with a magnetization along the [110] direction. \label{struct}}.
\end{figure}

As it is illustrated in Figs. \ref{struct}(a) and (b), the vanadium-phosphorous-oxide $\beta$-V$_2$OPO$_4$ compound crystalizes in body-centered-tetragonal (BCT) structure with a space group $I4_1amd$ (No. 141). Structural optimization obtains the calculated lattice constants which are $a=5.465$ {\AA} and $c=12.544$ {\AA}, in nice agreement with experimental values $a=5.362$ {\AA} and $c=12.378$ {\AA} \cite{Glaum1989}. The O atoms take two Wyckoff positions $4a$ (0.0, 0.0, 0.0) and $16h$ (0.00000, 0.23762, 0.43299). The P and V atoms are located at Wyckoff positions $4b$ (0.0, 0.0, 0.5) and $8c$ (0.250, 0.000 0.875), respectively. The BCT BZ and corresponding (001) surface BZ are shown in Fig. \ref{struct}(c), in which high-symmetry points are marked.

\begin{figure}
	\centering
	\includegraphics[scale=0.4]{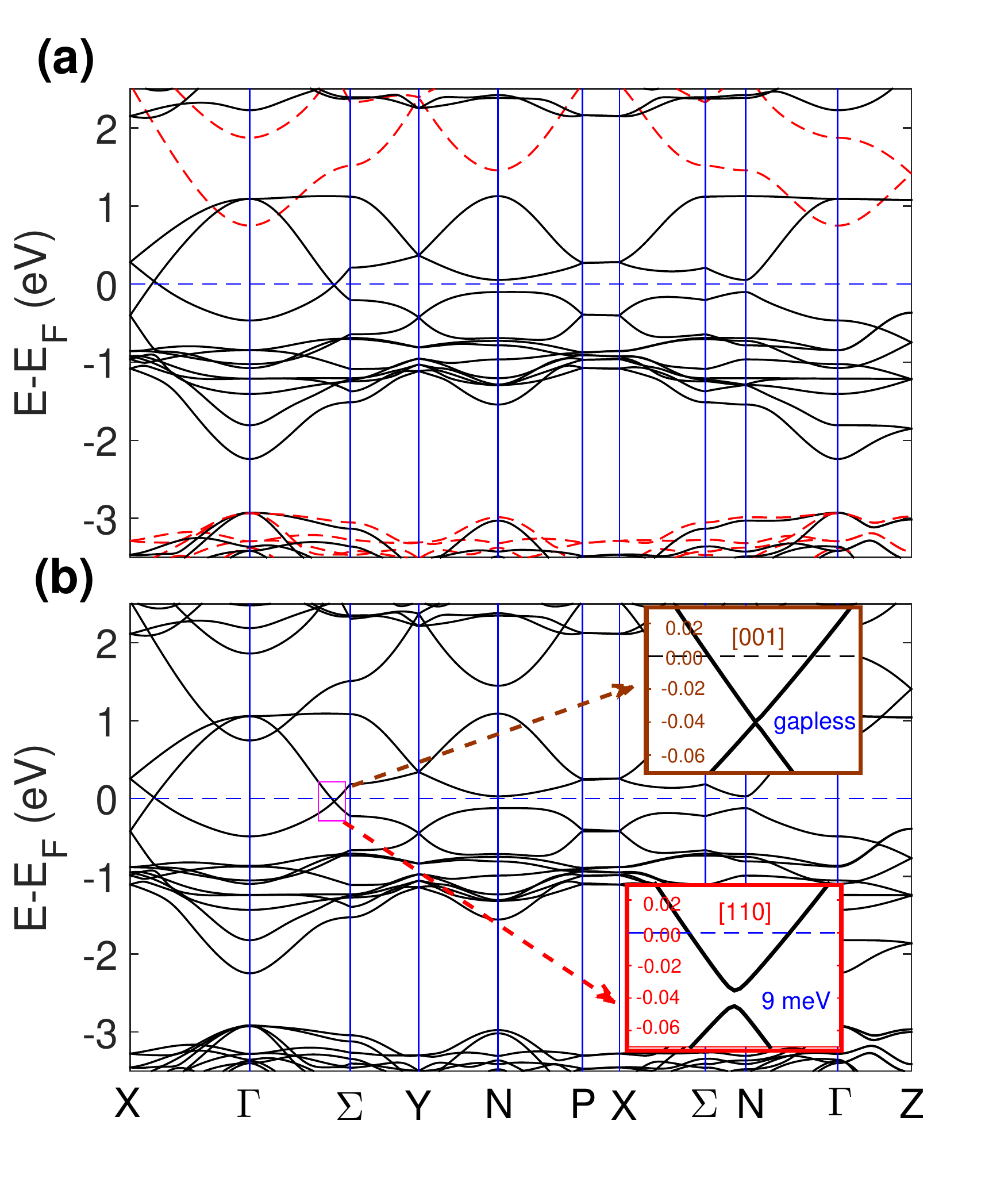}
	\caption{(a) The band structures of $\beta$-V$_2$OPO$_4$ along high-symmetry lines without spin-orbital coupling (SOC), and the majority and minority spin bands are denoted by the solid (black) and dashed (red) lines, respectively. (b) The SOC band structures. The upper and lower insets respectively represent the bands with the [001] and [110] magnetism in $\Gamma$-$\Sigma$ direction. \label{band}}.
\end{figure}

Our results confirm the FM groundstate of $\beta$-V$_2$OPO$_4$, which is about 86 meV lower than the antiferromagnetic state per unit cell, with magnetic moment $\sim$2.5 $\mu_{B}$ per V atom. The electronic band structures in the absence of SOC [see Fig. \ref{band}(a)] show that the spins and orbitals are independent and two spin channels are decoupled around the Fermi level. These two spin channels present different electronic states, i.e. a $\sim$3.68 eV band gap of the minority spin states and the semimetallic features of majority spin states, indicating the half-metallic properties of the system. The band crossings reveals around $\Gamma$ point in the $k_z = 0$ plane, giving rise to a symmetry protected nodal-line loop. These two crossing bands belong to the states that has opposite eigenvalues $\pm$1 of mirror reflection symmetry operation $M_z$, respectively, which protects the nodal ring in $k_x$-$k_y$ plane with $k_z = 0$ \cite{Weng2015}. The spin-polarized nodal ring shows a tiny dispersion, the maximum and minimum of which are in the $\Gamma$-X and $\Gamma$-$\Sigma$ directions, respectively, i.e. $\sim$12 meV higher and $\sim$36 meV lower than the Fermi level $E_F$.

%Our results confirm the FM ground state in $\beta$-V$_2$OPO$_4$ with magnetic moment $\sim$2.5 $\mu_{B}$  per V atom. The energy of FM state is about 86 meV lower than the antiferromagnetic (AFM) state per unit cell according to GGA+U calculations. The electronic band structures in the absence of SOC are shown in Fig. \ref{band}(a). In the spin-polarized calculation without SOC, the spin and orbitals are independent and two spin channels are decoupled. Here, we can see a $\sim$3.68 eV band gap of the minority spin states, while the majority spin states show semimetallic features. The results suggest that this system exhibits good half-metallic properties. In the majority spin states, two band crossings occur along $\Gamma$-X and $\Gamma$-$\Sigma$ directions. In fact, the band crossings may happen around $\Gamma$ point in the $k_z = 0$ plane, giving rise to a nodal line. The two crossing bands belong to opposite mirror eigenvalues $\pm$1 with respect to mirror reflection symmetry $M_z$, which protects the nodal ring in $k_z = 0$ plane \cite{Weng2015}. The spin-polarized nodal line of majority spin states shows a tiny dispersion. The maximum and minimum of this dispersion are in the $\Gamma$-X and $\Gamma$-$\Sigma$ directions, respectively, which are $\sim$12 meV higher and $\sim$36 meV lower than the Fermi level $E_F$.

As we report in Fig. \ref{band}(b), the SOC little influences on the band structures in which the half-metallic ferromagnetism remains. The spontaneous magnetization direction is determined by studying the total energy of system with magnetization along different high symmetry axis. The [001] axis is found to be the energetically most favorable magnetization direction. It is worth to mention that, results show very small energy differences, below 0.3 meV, among all magnetic configurations. This implies that the switching between each configurations by external magnetic field would be easy.  In the presence of SOC, the magnetic symmetry is dependent on the direction of magnetization. We will respectively present the topological features of the two typical [001] and [110] magnetisms in the following.

%In the presence of SOC, we can see that the SOC little influences on the band structures as shown Fig. \ref{band}(b). The half-metallic ferromagnetism is also remained due to the small SOC strengths of V atom. In order to determine the spontaneous magnetization direction, we perform first-principles total energy calculations with a series of magnetization directions along possibly high symmetry axis. The [001] axis is found to be the energetically most favorable magnetization direction. However, our calculations indicate that the energy differences among all magnetic configurations are smaller than 0.3 meV.  Within SOC, the magnetic symmetry is dependent on the direction of spontaneous magnetization, so we will respectively present the topological features of the two typical [001] and [110] magnetisms in the following.

\begin{figure}
	\centering
	\includegraphics[scale=0.06]{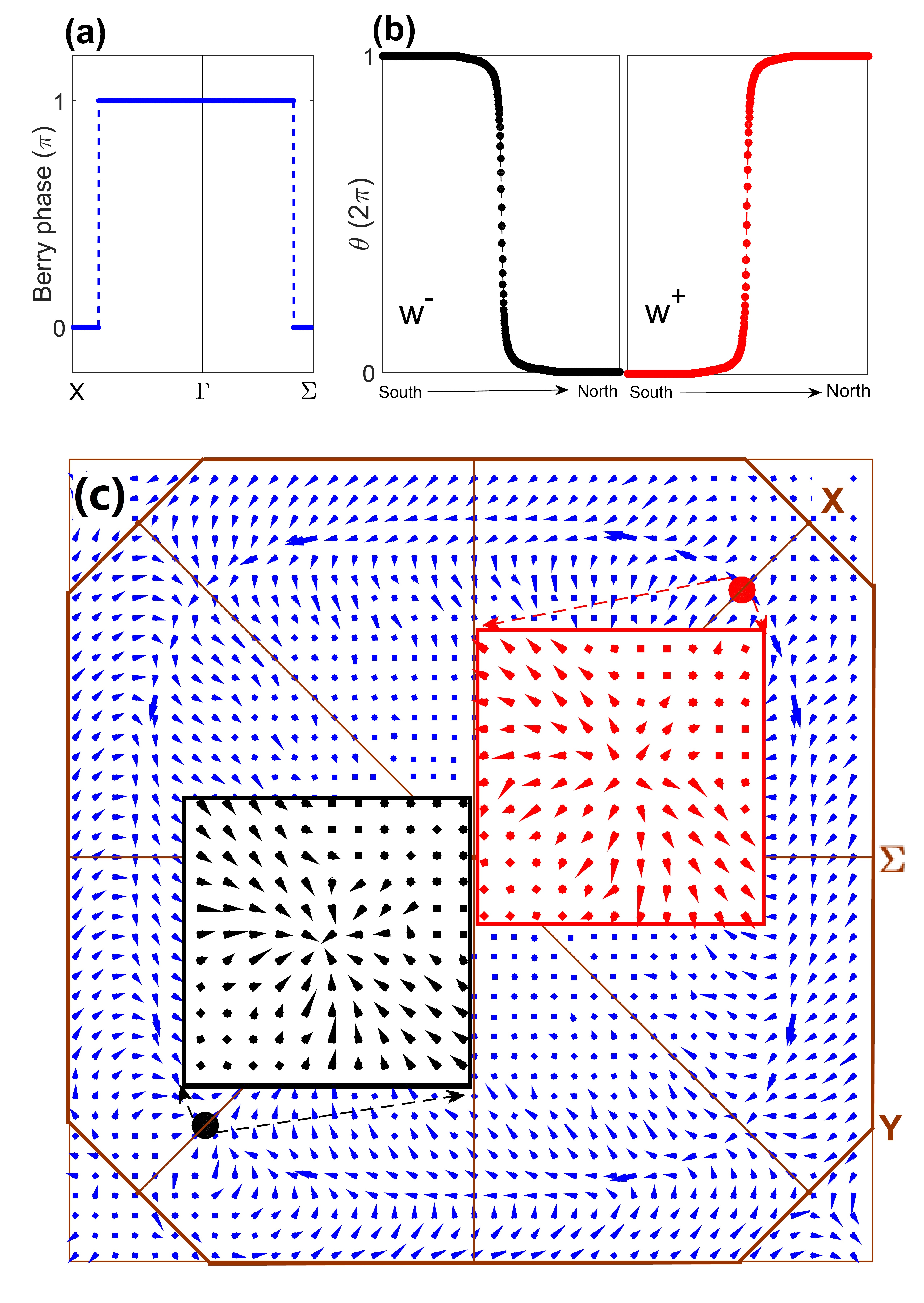}
	\caption{(a) With magnetization direction along [001] axis, variation of the Berry phase along high-symmetry lines in $k_z = 0$ plane. With magnetization direction along [110] axis, (b) the evolution of average position of Wannier centers obtained by the Wilson-loop method applied on a sphere that encloses on a Weyl point. The average Wannier center shifts downwards from south to north, indicating to negative chirality $W^{-}$. The average Wannier center shifts upwards from south to north, indicating to positive chirality $W^{+}$. (c) The distribution of Berry curvature for the $k_z = 0$ plane. The red and black regions denote the Weyl points with negative and positive chirality, respectively. \label{berry}}.
\end{figure}

%When the FM magnetization parallel to the [001] direction, the crossings along $\Gamma$-X (or [110]) and $\Gamma$-$\Sigma$ (or [100]) directions are gapless as shown in Fig. \ref{band}(b), and the topological nodal loop is still remained.  When all spins are oriented along [011](or $z$) direction, the $C_{4h}$ subgroup of $D_{4h}$ is just the fourfold rotation group $C_{4}^{z}$ with respect to $z$ coordinate axis tensored by the inversion
%$I$, namely $C_{4}^{z}\otimes I$. The magnetic symmetry contains eight irreducible representations: inversion $I$, fourfold rotation $C_{4}^{z}$, the product of time reversal and twofold rotations of $C_2$ symmetry axes [100], [010], [110], $1\bar{1}0$, and the product ($I\cdot C_{2}^z$) of inversion and rotation $C_{2}^z$. The group element $I\cdot C_{2}^z$ is just the mirror reflection symmetry with respect to the $xy$ plane, which can protect the topological nodal ring remains in $k_{z}=0$ plane with respect to SOC \cite{Weng2015, Wang2016prl1}.  The topological invariant of the nodal ring can be viewed as the variation of the quantized Berry phase \cite{Vanderbilt1993}, which is related to the change at the end of the one-dimensional system along a line across the ring in  $k_{z}=0$ plane. As shown in Fig. \ref{berry}(a), the Berry phase of $\beta$-V$_2$OPO$_4$  shows the jump across the ring, further confirming the topological features of the nodal ring perpendicular to the [001] magnetization direction.

 When magnetization is polarized along [001] direction, the system reduces to point group $C_{4h}$, the subgroup of $D_{4h}$, in which the fourfold rotation $C_{4}^{z}$ is tensored by the inversion $I$, namely $C_{4}^{z}\otimes I$. This magnetic group contains eight irreducible symmetry operators: inversion $I$, fourfold rotation $C_{4}^{z}$, the product of time reversal $T$, twofold rotations of $C_2$ symmetry axes [100], [010], [110], [$1\bar{1}0$], and the product ($IC_{2}^z$) of inversion and rotation $C_{2}^z$. The group element $I C_{2}^z$ is equivalent to the mirror reflection symmetry corresponding to the $x-y$ plane, which can protect the existence of gapless nodal ring in  $k_{z}=0$ plane with respect to SOC \cite{Weng2015, Wang2016prl1} [see Fig. \ref{band}(b)]. The topological invariant of the nodal ring can be viewed as the variation of the quantized Berry phase with respect to the mirror plane \cite{Vanderbilt1993}, which is related to the change at the end of the one-dimensional system along a line across the ring in  $k_{z}=0$ plane. As shown in Fig. \ref{berry}(a), the Berry phase of $\beta$-V$_2$OPO$_4$ shows the jump across the ring, further confirming the topological features of the nodal ring perpendicular to the [001] magnetization direction.

Our calculations suggest that the magnetization along other directions is energetically very close to the [001]. When the magnetization is deviated from [001] direction, the mirror reflection symmetry is broken.  Here, we take the case of [110] magnetization as an example since the symmetry analysis for the rest cases are essentially the same. The group elements of the corresponding magnetic space group $C_{2h}$ remains: $I$, $C_2^{110}$ and $TC_{2}^z$, $TC_2^{1\bar{1}0}$. The vanishing of mirror reflection symmetry makes nodal line gapped. However, the anti-unitary symmetry $TC_{2}^z$ allows the existence of Weyl points in  $k_{z}=0$ plane. A pair of Weyl points protected by the $C_2^{110}$ rotation is present on $k_x$=$k_y$ axis.

We also calculate the parities of inversion eigenvalues at time reversal invariant momenta (TRIM) points. The product of the occupied bands running over all TRIM points is -1, confirming the presence of odd number of pairs of Weyl points \cite{Hughes}.  As shown in Fig. \ref{band}(b), the two crossing bands along $\Gamma$-X (or [110]) belong to eigenvalues $\pm i$ of $C_2^{110}$, respectively. The chirality of Weyl point can be determined by the evolution of the average position of Wannier centers, and the Wilson-loop method applied on a sphere around a Weyl point is used \cite{Yu2011,Soluyanov2011} [see Fig. \ref{berry}(b)].  The Weyl point with Chern number $C=+1$ is located at (0.46{\AA}$^{-1}$, 0.46{\AA}$^{-1}$, 0) in momentum space, while the Weyl point with Chern number $C=-1$ related by $I$ symmetry located at same axis [see Fig. \ref{struct}(c)]. The Weyl points only existed on $k_x$=$k_y$ axis can be further verified by the Berry curvature. As shown in Fig. \ref{berry}(c), the Weyl points with positive and negative chirality are regarded as the "source" and "sink" of Berry curvature in momentum space.

As it is discussed above, our results indicate the existence of either the topological NLs or WSMs in FM $\beta$-V$_2$OPO$_4$ compound, depending on the magnetization direction. The NL features are protected by the mirror reflection symmetry in the case of [001] magnetization, while the WSM features can arise with magnetization direction along other high symmetry axis. The Weyl points would be protected by $C_2^{m}$ rotational symmetry, where $m$ represents a symmetry axis in $x-y$ plane. In the cases of which only a pair of Weyl points, the minimum number of Weyl points in condensed matter systems, with large separation in momentum. Considering the energy of Weyl points in FM $\beta$-V$_2$OPO$_4$ is very close to Fermi level, e.g. $\thicksim$12 meV when [110] magnetization, we would suggest $\beta$-VOPO$_4$ compound to be a excellent experimental candidates for the observation of the nontrivial properties in FM WSMs.

\begin{figure}
	\centering
	\includegraphics[scale=0.07]{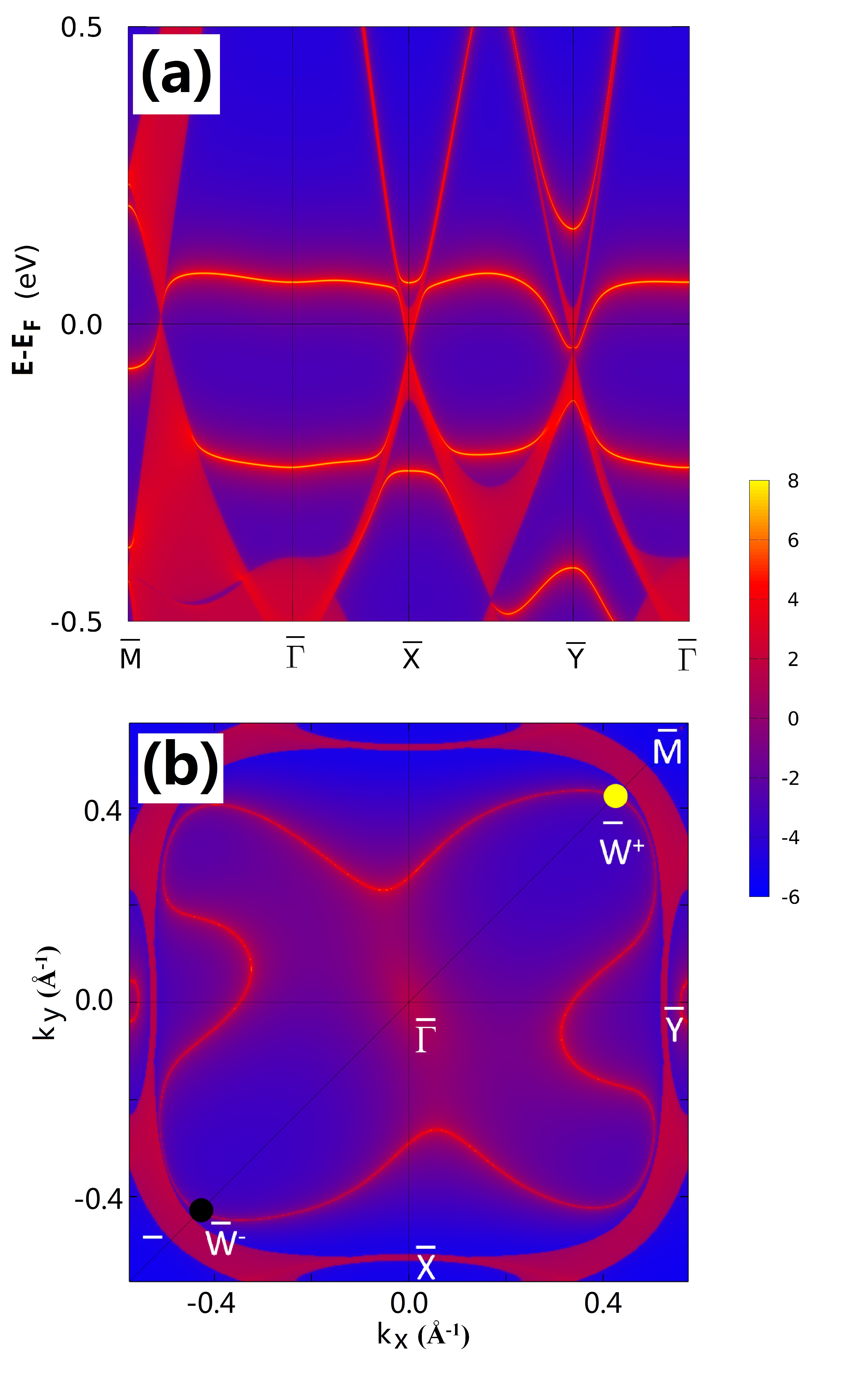}
	\caption{The surface states and Fermi surfaces. (a) The surfaces projected on (001) surfaces. (b) The corresponding Fermi surfaces on (001) surfaces. The projected
Fermi surface is calculated with a chemical potential of 60 meV. \label{surf}}.
\end{figure}

%One hallmark of nontrivial semimetals is the existence of topologically protected surface state, which arises from the inversion of TRIM parities. The Weyl points result from the nodal line via transforming the magnetization axis, so we present the surface and Fermi arc states in the [110] magnetic configuration. To obtain the surface states, we constructed a tight-binding (TB) Hamiltonian with basis of maximally localized Wannier \cite{Marzari2012,Mostofi2008}. In Wannier representation, the Green's function method \cite{Sancho1985} is employed.  The calculated local density of states (LDOS) and Fermi surface projected on (001) surface face are shown in Figs. \ref{surf}(a) and (b), respectively. The surface states are clearly observed in Fig. \ref{surf}(a). On the (001) surface, the magnetic symmetry $T\cdot C_{2}$ leads to different behavior for the surface bands around the $\bar{X}$ and $\bar{Y}$.  Since the SOC only lead to small gap along the nodal line, some trivially residual bands would project on any surface of this compound. However, the Fermi arc states across 60 meV above the Fermi level are still clear in Fig. \ref{surf}(b).

One hallmark of nontrivial semimetals is the existence of topologically protected surface state, which arises from the inversion of TRIM parities. In the WSM states, topological surface bands connect the valence and conduction bands. The drumhead surface states of NLSs will appear when the gap of bulk band is closed forming a nodal ring. Since the gap size of WSM state is very small, which make only tiny difference of surface states between NL and WSM phases. Here, we only present the surface states and Fermi arcs in the [110] magnetic configuration. To obtain the surface states, we constructed a tight-binding (TB) Hamiltonian with basis of maximally localized Wannier functions \cite{Marzari2012,Mostofi2008} in which the Green's function method \cite{Sancho1985} is employed.  The calculated local density of states (LDOS) and Fermi surface projected on (001) surface are shown in Figs. \ref{surf}(a) and (b), respectively. The surface states are clearly observed in Fig. \ref{surf}(a). On the (001) surface, the anti-unitary magnetic symmetry $TC_{2}$ leads to different behavior for the surface bands around the $\bar{X}$ and $\bar{Y}$.  Although some trivially residual bands would project on (001) surface of this compound, the Fermi arc states are quite clear as it is shown in Fig. \ref{surf}(b).

In conclusion, using first-principles calculations, we proposed that FM vanadium-phosphorous-oxide $\beta$-V$_2$OPO$_4$ can possess the nontrivial properties of TSMs, either NLs or WSMs, that can be switched between each other by different magnetization directions. When the spin is polarized along [001] direction, system belongs to $D_{4h}(C_{4h})$ point group and present the gapless nodal ring, protected by the mirror reflection symmetry, close to the Fermi level at $k_x - k_y$ plane with $k_z=0$ in momentum space. When the magnetization directions deviate from [001], the nodal ring will degenerate to a pair of Weyl points due to the vanishing of mirror reflection symmetry. These two Weyl points, protected by twofold rotational symmetry along magnetization directions, lie in $k_x - k_y$ plane with $k_z=0$, which are largely separated in momentum. All the non-trivial band-crossing points are very close to the Fermi level. The topology of the system is confirmed by the existence of non-trivial surface states. Since $\beta$-V$_2$OPO$_4$ has been fabricated for years \cite{Glaum1989}, our proposition provides a realistic and promising candidate for the investigation of magnetic TSMs in experiments, particularly towards the realization of quantum anomalous Hall effect condensed matter systems.

%In conclusion, using first calculations, we have found that FM vanadium-phosphorous-oxide $\beta$-V$_2$OPO$_4$ can possess the topologically nontrivial properties of WSMs. The current compound, in which there are only two Weyl points largely separated in momentum space and very close to Fermi level, reveals a nice candidate with the minimum number of Weyl points in a condensed matter system.  In the magnetic system with spatial-inversion, the topological features are confirmed by the parities of inversion eigenvalues at TRIM points. The Weyl points arise from the nodal ring gapped due to the mirror reflection symmetry broken in the presence of SOC. It is demonstrated that the Weyl points are protected by the twofold rotational symmetry $C_2^m$ along the magnetization direction. Interestingly, along a special magnetization direction along [001] axis, $\beta$-V$_2$OPO$_4$ can possess the node-line semimetallic features even though the SOC is included. Surface band calculations reveal non-trivial Fermi arcs states as well. This proposition provides a realistic and promising platform for the investigation of magnetic Weyl phenomena in experiments, particularly towards the realization of quantum anomalous Hall effect in Weyl semimetals.

\emph{Acknowledgments--}
This work is supported by the National Natural Science
Foundation of China (NSFC, Grant Nos.11204185, 11304403,
11334003 and 11404159).
~~~\\
\\
\textbf{AUTHOR INFORMATION}\\
$^*$\textbf{Equal Contributions:}\\
Y. J. Jin and R. Wang contributed equally to this work.\\
$^{\dag}$\textbf{Corresponding author:}\\
xuh@sustc.edu.cn (H.X.)\\

\clearpage
\newpage

\setcounter{figure}{0}
\makeatletter

\makeatother
\renewcommand{\thefigure}{S\arabic{figure}}
\renewcommand{\thetable}{S\Roman{table}}
\renewcommand{\theequation}{S\arabic{equation}}

\begin{center}
	\textbf{
		\large{Supplemental Material for}}
	\vspace{0.2cm}
	
	\textbf{
		\large{
			``Topological Semimetals in Ferromagnetic Vanadium-Phosphorous-Oxide $\beta$-V$_2$OPO$_4$ Compound" }
	}
\end{center}

\section{Two-band $2\times 2$  $k\cdot p$  Hamiltonian}

In the absence of SOC, a node-line around the $\Gamma$ point in $k_z = 0$ plane is present. Generally, the nodal ring around the $\Gamma$ point can be modeled by a two-band $k\cdot p$ theory, and the Hamiltonian is
\begin{equation}\label{H}
H=\sum_{i=x,y,z}d_{i}(\mathbf{k})\sigma_{i},
\end{equation}

where $d_{i}(\mathbf{k})$ are real functions and $\mathbf{k}=(k_x,k_y,k_z)$ are three components of the momentum $\mathbf{k}$ relative to the $\Gamma$ point. In Eq. (\ref{H}), we have ignored the kinetic term proportional to the identity matrix, since it is irrelevant in studying the band crossing. The Pauli matrix $\sigma_{i}$ are
\begin{equation}
\sigma_{x}=\left(
             \begin{array}{cc}
               0 & 1 \\
               1 & 0 \\
             \end{array}
           \right)
,
\sigma_{y}=\left(
             \begin{array}{cc}
               0 & -i \\
               i & 0 \\
             \end{array}
           \right),
 \sigma_{z}=\left(
             \begin{array}{cc}
               1 & 0 \\
               0 & -1 \\
             \end{array}
           \right)
\end{equation}

Within SOC, two spin channels couple together and symmetries can decreasing depending on the direction of the spontaneous magnetization. In spin representation, an any three dimensional (3D) rotation $R(\alpha, \beta, \gamma)$ has to correspond a two-dimensional (2D) unitary matrix $u(\alpha, \beta, \gamma)$ as \begin{equation}\label{spinu}
u(\alpha, \beta, \gamma)=\left(
                           \begin{array}{cc}
                             e^{-i\frac{\alpha+\gamma}{2}}\cos \frac{\beta}{2} & -e^{-i\frac{\alpha-\gamma}{2}}\sin \frac{\beta}{2} \\
                             e^{i\frac{\alpha-\gamma}{2}}\sin \frac{\beta}{2} & e^{i\frac{\alpha+\gamma}{2}}\cos \frac{\beta}{2} \\
                           \end{array}
                         \right),
\end{equation}

where $\alpha$, $\beta$, $\gamma$ are Euler angle of 3D rotation.

\section{The nodal ring depending on magnetic space group $D_{4h}(C_{4h})$ in the presence of spin-orbital-coupling}
When all spins are oriented along [011](or $z$) direction, the $C_{4h}$ subgroup of $D_{4h}$ is just the fourfold rotation group $C_{4}^{z}$ with respect to $z$ coordinate axis tensored by the inversion
$I$, namely $C_{4}^{z}\otimes I$. With [100] magnetization, the group elements of the corresponding magnetic space group $D_{4h}(C_4h)$ remains: $I$, $C_{4x}$, $C_{2}^{z} \cdot I$, $C_2^{010} \cdot T$, and $C_2^{100}\cdot T$. It is important note that the product $C_{2}^{z} \cdot I$ of twofold rotation $C_{2}^{z}$ and inversion $I$ is mirror-reflection symmetry $M_z$ with respect to (001) plane. The matrix form of $M_z$ in spin representation is
\begin{equation}
C_{2}^{z} \cdot I = \left(
             \begin{array}{cc}
               e^{i\frac{\pi}{2}} & 0 \\
               0 & e^{-i\frac{\pi}{2}} \\
             \end{array}
           \right)=i\sigma_{z}.
\end{equation}
The mirror reflection indicates the Hamiltonian in $xy$ [or (001)] plane as
\begin{equation}
H(k_x,k_y,0)=\sigma_z H(k_x,k_y,0)\sigma_{z}
\end{equation}
or
\begin{equation}
\begin{split}
&d_x(k_x,k_y,0)=-d_x(k_x,k_y,0)\equiv 0,\\
&d_y(k_x,k_y,0)=-d_y(k_x,k_y,0)\equiv 0,\\
\end{split}
\end{equation}
The energy dispersion of the two-band Hamiltonian with  $C_{2}^{z} \cdot I$ symmetry is
\begin{equation}
E=\pm |d_z(k_x,k_y,0)|.
\end{equation}
Generically, the band crossing means $d_z(k_x,k_y,0)=0$, which has codimension one, i.e., a nodal loop solution in $xy$ plane.

\section{The Weyl points depending on magnetic space group $D_{4h}(C_{2h})$ in the presence of spin-orbital-coupling}
We show that the Weyl points can arise with magnetization direction along high symmetry axis, as long as the magnetic space group $C_{2h}$ symmetry remains. The Weyl points are protected by $C_2^{m}$ rotation, where $m$ represents a symmetry axis in $xy$ [or (001)] plane. As a example, we mainly show the case of magnetization along [100] axis.

With [100] magnetization, the group elements of the corresponding magnetic space group $D_{4h}(C_2h)$ remains: $I$, $C_{2}^{z} \cdot T$, $C_2^{010} \cdot T$, and $C_2^{100}$. We now analyze the group elements on different axes to find out if Weyl points can arise generically on those axes in (001) plane. Firstly, we prove the product $C_{2}^z \cdot T$ of time reversal $T=-i\sigma_{y} K$ and rotation $C_{2}^{z}$ allowing for the existence of Weyl points in the (110) plane. The anti-unitary $C_{2}^z \cdot T$ with matrix representation is
\begin{equation}
\begin{split}
C_{2}^z \cdot T &= i\left(
             \begin{array}{cc}
               e^{-i\frac{\pi}{2}} & 0 \\
               0 & e^{i\frac{\pi}{2}} \\
             \end{array}
           \right)\left(
             \begin{array}{cc}
               0 & -i \\
               i & 0 \\
             \end{array}
           \right) K \\
           &= i\sigma_{x}K,
\end{split}
\end{equation}
with complex conjugation $K$. The anti-unitary $C_{2}^z \cdot T$ requires the commutation relation as
\begin{equation}
[H, C_{2}^z \cdot T]=[H, i\sigma_{x}K]=0,
\end{equation}
i.e.,
\begin{equation}
\begin{split}
& d_{x} (k_x,k_y,0)[\sigma_{x},i\sigma_{x}K]+d_{y} (k_x,k_y,0)[\sigma_{y},i\sigma_{x}K]\\
&+d_{z}(k_x,k_y,0)[\sigma_{z},i\sigma_{x}K]=0,
\end{split}
\end{equation}
which gives
\begin{equation}
d_z (k_x,k_y,k_z)\equiv 0.
\end{equation}
Now the Hamiltonian in (001) plane is
\begin{equation}
H=d_x(k_x,k_y,0)\sigma_{x}+d_y(k_x,k_y,0)\sigma_{y}.
\end{equation}
It is noted that the Hamiltonian contains two parameters and two momenta. Hence the crossing points can generically exist in the (001) plane.

In the following, we show that the Weyl points can arise from the unitary element $C_{2}^{100}$ with [100] magnetization axis. This symmetry requires the constraints:
\begin{equation}\label{H100}
[H,C_{2}^{100}]=0.
\end{equation}

In spin representation, the unitary matrix $C_{2}^{100}$ can be obtained Eq. (\ref{spinu}), as
\begin{equation}
C_{2}^{100}=\left(
\begin{array}{cc}
               0 & i \\
               i & 0 \\
             \end{array}
           \right)=i\sigma_{x}.
\end{equation}
Eq. (\ref{H100}) gives
\begin{equation}
H(k_x,k_y,0)=\sigma_x H(k_x,-k_y,0)\sigma_{x}
\end{equation}
or
\begin{equation}\label{H100D}
\begin{split}
&d_x(k_x,k_y,0)=d_x(k_x,-k_y,0),\\
&d_y(k_x,k_y,0)=-d_y(k_x,-k_y,0).
\end{split}
\end{equation}

In $k_y =0$ axis, i.e. [100] direction, Eq. (\ref{H100D}) must requires
\begin{equation}
d_y(k_x,0,0)\equiv 0.
\end{equation}
The energy dispersion of the two-band Hamiltonian after considering $C_{2}^{100}$ symmetry is
\begin{equation}
\begin{split}
E&=\pm\sqrt{d_x(k_x,k_y,0)^2+d_y(k_x,k_y,0)^2}\\
 &=\pm |d_x (k_x,0,0)|.
\end{split}
\end{equation}
In low-energy case, we have
\begin{equation}
d_x(k_x,0,0)=A+B k_x +C k_x^2+\ldots,
\end{equation}
and then the zero energy mode can require a pair of Weyl points locate at $k_x=\pm k_x^c$ when $B=0$ and $AC<0$. Two bands with eigenvalues $\pm i$ of $C_2^{100}$ can cross on this axis. Along other direction with $k_y = a k_{x}$ through $\Gamma$ point on the $xy$ plane, the Hamiltonian contains two function $d_x$ and $d_y$, but one momentum $k_x$, indicating that there are no Weyl points on the $k_y = a k_{x}$ line. Hence, the crossing would splits away from [100] axis. Similarly, when magnetization is along [110], [010], and $[1\bar{1}0]$ directions, the Weyl points also arise due to  $C_2^{m}$ rotation of magnetic space group $D_{4h}(C_{2h})$. The Weyl points are only allowed in $m$ axis.

\end{document}